\documentclass[authoryear]{elsarticle}
\usepackage{natbib,amssymb}
\usepackage{amsmath}
\usepackage{amsfonts}
\usepackage{graphicx}
\usepackage{multicol}
\usepackage{multirow}
\newcommand{\D}{{\mathrm{d}}}

\newtheorem{proposition}{Proposition}
\newtheorem{definition}{Definition}

\begin{document}
\begin{frontmatter}

\title{Evolution of adaptation mechanisms: adaptation energy, stress, and oscillating death}
\author{Alexander N. Gorban}
 \ead{ag153@le.ac.uk}
\author{Tatiana A. Tyukina}
\ead{tt51@leicester.ac.uk}
\address{University of Leicester, Leicester, LE1 7RH, UK}
\author{Elena V. Smirnova}
\ead{seleval2008@yandex.ru}
\author{Lyudmila I. Pokidysheva}
\ead{pok50gm@gmail.com}
\address{Siberian Federal University, Krasnoyarsk, 660041, Russia}

\begin{abstract}
In 1938, H. Selye proposed the notion of adaptation energy and published
``Experimental evidence supporting the conception of adaptation energy''.  Adaptation of an animal to different factors appears as the spending of one resource.  Adaptation  energy is a
hypothetical extensive quantity spent for adaptation.
This term causes much debate when one takes it literally, as a
physical quantity, i.e. a sort of energy. The controversial points of view impede the systematic use of the notion of adaptation energy despite experimental evidence.
Nevertheless, the response to many harmful factors often has general non-specific form and we suggest that the
mechanisms of physiological adaptation admit a very general and nonspecific description.

We aim to demonstrate that Selye's adaptation energy is the cornerstone of the top-down approach to modelling of
non-specific adaptation processes. We analyse Selye's axioms of adaptation energy
together with Goldstone's modifications and propose a series of models for interpretation
of these axioms. {\em Adaptation energy is considered as an internal coordinate on the
`dominant path' in the model of adaptation}. The phenomena of `oscillating death' and
`oscillating remission' are predicted on the base of the dynamical models of adaptation.
Natural selection plays a key role in the evolution of mechanisms of
physiological adaptation. We use the fitness optimization approach to study of  the  distribution of resources for neutralization of harmful factors, during adaptation to a multifactor environment, and analyse the optimal strategies for different systems of factors.
\end{abstract}

\end{frontmatter}

\section{Introduction}

\cite{SelyeAEN} introduced the notion of adaptation energy as the universal currency for
adaptation. He published  ``Experimental evidence supporting the conception of
adaptation energy'' \citep{SelyeAE1}: adaptation of an animal to different factors
(sequentially) looks like spending of one resource, and the animal dies when this
resource is exhausted.

The term `adaptation {\em energy}' contains an attractive metaphor: there is a
hypothetical extensive variable which is a resource spent for adaptation. At the same
time, this term causes much debate when one takes it literally, as a
physical quantity, i.e. as a sort of energy, and asks to demonstrate the physical nature of
this `energy'. Such discussions impede the systematic use of the notion of adaptation
energy even by some of  Selye's followers. For example, in the modern ``Encyclopedia of
Stress'' we read: ``As for adaptation energy, Selye was never able to measure it...''
\citep{AEencicl}. Nevertheless, this notion is proved to be useful in the analysis of
adaptation \citep{BreznitzAEappl,SchkadeOccAdAE2003}.

Without any doubt, adaptation energy is not a sort of physical energy. Moreover, Selye
definitely measured the adaptation energy: the natural measure of it is the intensity and
length of the given stress from which adaptation can defend the organism before {\em
adaptability} is exhausted. According to \cite{SelyeAE1}, ``during adaptation to a
certain stimulus the resistance to other stimuli decreases''. In particular, he
demonstrated that  ``rats pretreated with a certain agent will resist such doses of this
agent which would be fatal for not pretreated controls. At the same time, their
resistance to toxic doses of agents other than the been adapted decreases below the
initial value.''

These findings were tentatively interpreted using the assumption that the resistance of the
organism to various damaging stimuli depends on its adaptability. This adaptability
depends upon adaptation energy of which the organism possesses only a
limited amount, so that if it is used for adaptation to a certain stimuli, it will
necessarily decrease.

\cite{SelyeAE1} concluded that ``adaptation to any stimulus is always acquired at a cost,
namely, at the cost of adaptation energy.'' No other definition of adaptation energy was
given. This is just a resource of adaptability, which is spent in all adaptation
processes.  The economical metaphors used by Selye, `cost' and `spending', were also
seminal and their use was continued in many works. For example, \cite{Goldstone1952}
considered adaptation energy as a ``capital reserve of adaptation'' and death as ``a
bankruptcy in non-specific adaptation energy''.

The economical analogy is useful in physiology and ecology for analysis of interaction of
different factors. \cite{GorSmiCorAd1st} analysed interaction of factors in human
physiology and demonstrated that adaptation makes the limiting factors equally important.
These results underly the method of correlation adaptometry, that measures the level of
adaptation load on a system and allows us to estimate health in groups of
healthy people \citep{Sedov}. For plants, the economical metaphor was elaborated by
\cite{BloomChapinMooney1985} and developed further by \cite{ChapinSchulzeMooney1990}.
They also merged the optimality and the limiting approach and used the notion of
`exchange rate' for factors and resources. For more details and connections to
economical dynamics we refer to  \cite{GorbanSmiTyu2010}. For
systems of factors with different types of interaction (without limitation) adaptation
may lead to different results \citep{GorbanPokSmiTyu}. In particular, if there is {\em
synergy} between several harmful factors, then adaptation should make the influence of
different factors uneven and may completely exclude (compensate) some of them.

In order to understand why  we need the notion of adaptation energy in modelling of
physiology of adaptation, we have to discuss two basic approaches to modelling, {\em
bottom-up} and {\em top-down}.
\begin{itemize}
\item The bottom-up approach to modelling in physiology ties molecular and cellular
    properties to the macroscopic behaviour of tissues and the whole organism. Modern
    multiagent methods of modelling  account for elementary interactions, and provide analysis how the rules of elementary events affect the macroscopic dynamics. For
    example, \cite{Galle2009} demonstrate how the individual based models explain
    fundamental properties of the spatio-temporal organisation of various multi-cellular
    systems. However, such models may be too rich and detailed, and typically, different
    model assumptions  comply with known experimental results equally well. In order
    to develop reliable quantitative individual based models, additional experimental
    studies are required for identifying the details of the elementary events
    \citep{Galle2009}. We suspect that for the consistent and methodical bottom-up
    modelling, we will always need additional information for identification of the
    microscopic details.
\item Following the top-down approach, we start from very general integrative properties
    of the whole system and then add some details from the lower levels of organization,
    if necessary. It is much closer to the classical physiological approach. A properly
    elaborated top-down approach creates the background, the framework and the
    environment for the more detailed models. We suggest, without exaggeration,  that
    all  detailed models need the top-down background (like quantum mechanics, which cannot
    be understood without its classical limit). The top-down approach allows one to relate the
    modelling process directly to experimental data, and  to test the model with
    clinical data \citep{Hester2011}. Therefore, the language of the problem statement
    and the interpretation of the results is generated using the top-down approach.
\item To combine the advantages of the bottom-up and the top-down approaches, the {\em
    middle-out} approach was proposed \citep{Brenner1998,Kohl2010}. The main idea is to start not
    from the upper level but from the level which is ready for formalization. That is the
    level where the main mechanisms are known, and it is possible to develop an adequate mathematical model
    without essential extension of experimental and theoretical basis. Then we can move
    upward (to a more abstract integrative level) or downward (to more elementary details),
    if necessary. Following \cite{Noble2003} we
    suggest that ``reduction and integration are just two complementary sides of the same
    grand project: to unravel and understand the `Logic of Life'.''
\end{itemize}

\cite{SelyeAE1} and later \cite{Goldstone1952} used the notion of adaptation energy to
represent the typical dynamics of adaptation. In that sense, they prepared the theory of
adaptation for mathematical modelling. The adaptation energy is the most integrative
characteristic for the models of top level. In this work, we develop a hierarchy of
top-down models following Selye's findings and further developments.

We follow Selye's insight about adaptation energy and provide a `thermo\-dy\-na\-mic-like' theory of organism resilience that (just like classical thermodynamics) allows for economic metaphors (cost, bankruptcy) and, more importantly, is largely independent of a detailed mechanistic explanation of what is  `going on underneath'.

We avoid direct discussion of the question of whether the adaptation energy is a `biological reality', a `generalizing term' for a set of some specific (unknown) properties  of an organism that provide its adaptation, or  `just a metaphor' similar to `phlogiston' or `ether',  notions that were useful for description of some phenomena but had no actual physical meaning as substances.

Moreover, we insist that the sense of the notion of adaptation energy is completely described by its place in the system of models like the notion of mass in Newtonian mechanics is defined by its place in the differential equations of Newton's laws. Selye did not write the equation of the adaptation energy but his experiments and `axioms' have been very `mathematical'. He proved that (in some approximation) there is an extensive variable (adaptation resource)  which an organism spends for adaptation. This resource was measured by  the intensity and length of various stresses from which adaptation can defend the organism.

\section{`Axioms' of adaptation energy }

Selye, Goldstone and some other researchers formulated some of their discoveries and
working hypotheses as `axioms'. These axioms, despite being different from
mathematical axioms, are used for fixing and securing sense. Selye's axioms of Adaptation
Energy (AE) (following \cite{SchkadeOccAdAE2003}) are:

\begin{enumerate}
\item AE is a finite supply, presented at birth.
\item As a protective mechanism, there is some
    upper limit to the amount of AE that an individual can use at any discrete moment in
    time. It can be focused on one activity, or divided among  other activities designed
    to respond to multiple occupational challenges.
\item There is a threshold of AE activation
    that must be present to potentiate an occupational response.
\item AE is active at two
    levels of awareness: a primary level at which creating the response occurs at a high
    awareness level, with high usage of finite supply of adaptation energy; and a
    secondary level at which the response creation is being processing at a sub-awareness
    level, with a lower energy expenditure.
\end{enumerate}
\begin{figure}
\centering{
\includegraphics[width=0.9\textwidth]{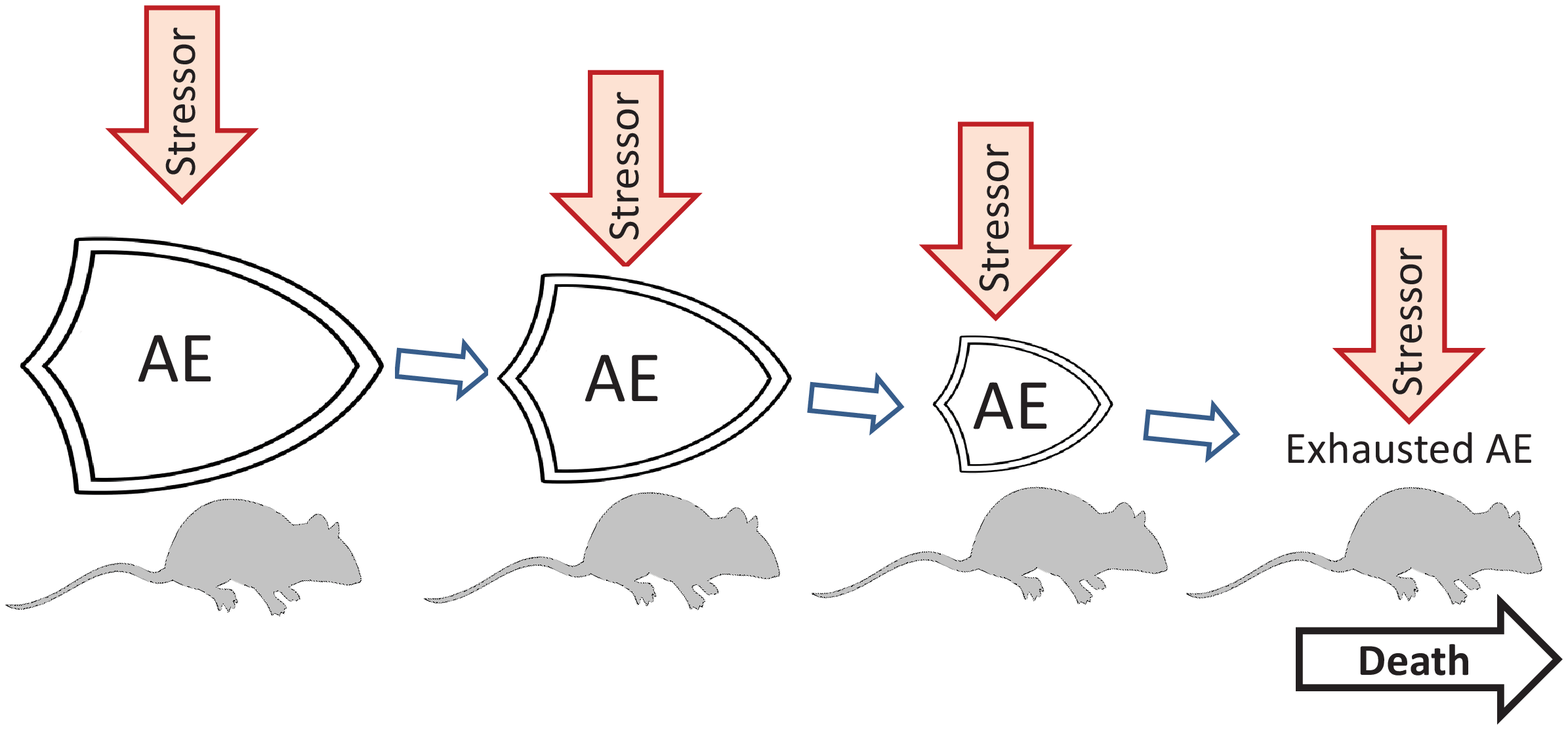}
}\caption{Schematic representation of Selye's axioms. The shield of adaptation spends AE
for protection from each stress. Finally, AE becomes exhausted, the animal cannot resist
stress and dies \label{schemeSel} (The rat silhouette is taken from Wikimedia commons,
File:Rat\_2.svg.)}
\end{figure}

Selye's Axioms 1-3 are illustrated by Fig.~\ref{schemeSel}.

\cite{Goldstone1952} proposed the concept of a constant production or income of AE
which may be stored (up to a limit), as a capital reserve of adaptation. He showed that
this concept best explains the clinical and Selye's own laboratory findings. According to
\cite{Goldstone1952}, it is possible that, had Selye's experimental animals been asked to
spend adaptation at a lesser rate (below their energy income), they might have been able to cope
successfully with their stressor indefinitely. The whole systems of adaptation reactions
to weaker factors was systematized by \cite{Garkavi1979}. On the basis of this system,
\cite{Garkavi1998} developed the activation therapy, which was applied in clinic,
aerospace and sport medicine.

\begin{figure}
\centering{
\includegraphics[width=0.9\textwidth]{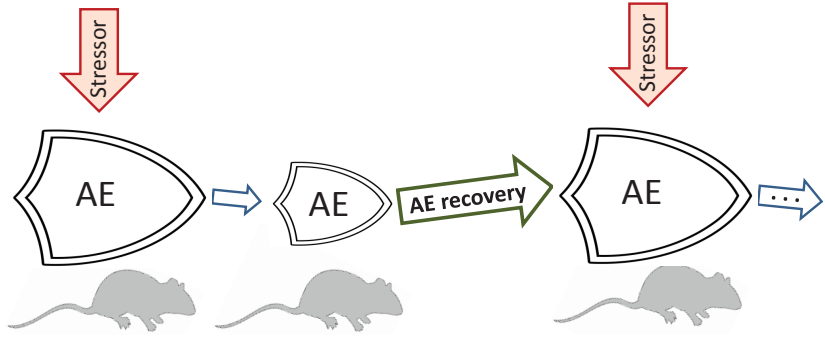}
}\caption{Schematic representation of Goldstone's modification of Selye's axioms: AE can
be recovered and adaptation shield may persist if there is enough time and reserve for
recovery. \label{schemeGold}}
\end{figure}

Goldstone's findings may be formulated as a modification of Selye's axiom~1. Their
difference from Selye's axiom~1 is illustrated by Fig.~\ref{schemeGold} (compare to
Fig.~\ref{schemeSel}). We call this modification Goldstone's axiom~1':
\begin{itemize}
\item AE can be created, though the income of this energy is slower in old
    age;
\item It can also be stored as Adaptation Capital, though the storage capacity has a
    fixed limit.
\item If an individual spends his AE faster than he creates it,
    he will have to draw on his capital reserve;
\item When this is exhausted he dies.
\end{itemize}

\section{Factor-resource basic model of adaptation}

Let us start from a simple (perhaps, the simplest) model with two phase variables, the
available free resource (AE) $r_0$ and the resource supplied for the stressor
neutralization, $r$. There are also four processes: degradation of the available
resource, degradation of the supplied resource, supply of the resource from the storage
$r_0$ to the allocated resource $r$, and production of the resource for further storage
($r_0$). The equations are:

\begin{equation}\label{simplest}
\begin{split}
&\frac{\D r}{\D t}=-k_d r+kr_0(f-r)h(f-r);\\
&\frac{\D r_0}{\D t}=-k_{d0}r_0 - kr_0(f-r)h(f-r)+k_{pr}(R_0-r_0),
\end{split}
\end{equation}
where
\begin{itemize}
\item $k_d r$ is the rate of degradation of resource supplied for the stressor
    neutralization, where $k_d$ is the corresponding rate constant;
\item $k_{d0}r_0$ is the rate of degradation of the stored resource, where $k_{d0}$ is the
    corresponding rate constant, we assume that $k_d\geq k_{d0}$;
\item $kr_0(f-r)h(f-r)$ is the rate of resource supply for the stressor neutralization, where $k$
    is the supply constant;
\item $h(f-r)$ is the Heaviside step function;
\item $k_{pr}(R_0-r_0)$ is the resource production rate, where $k_{pr}$ is the production rate
    constant.
\end{itemize}

Let us notice that:
\begin{itemize}
\item if $r_0 \geq R_0$ then $\D r_0/\D t\leq 0$,
\item  if $r_0=0$ then $\D r_0/\D t\geq 0$,
\item   if $r=0, r_0\geq 0$, then $\D r/\D t\geq 0$,
\item if $r=f$ then $\D r/\D t\leq     0$.
\end{itemize}
Therefore, the rectangle $D$ given by inequalities $0\leq r \leq f$, $0\leq r_0 \leq R_0$
is {\em positively invariant} with respect to system (\ref{simplest}): if the initial
values $(r(t_0),r_0(t_0))\in D$ for some time moment $t_0$ then the solution
$(r(t),r_0(t))\in D$ for $t>t_0$.

\begin{figure}
\centering{
\includegraphics[width=0.5\textwidth]{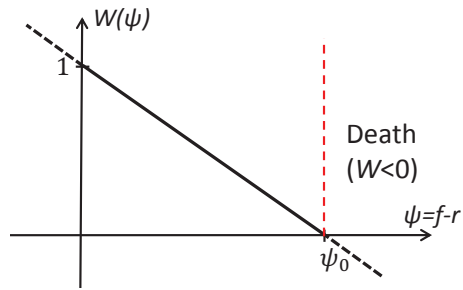}
}\caption{The fitness function for system (\ref{simple}). $\psi_0$ is the critical value
of stressor's intensity. If $f\leq \psi_0$ then life is possible without adaptation: for zero AE supply $W$ remains positive. \label{Fittness}}
\end{figure}

For large $f$ there exist a stable steady state in $D$ with
$$r_0\approx \frac{k_{pr}R_0}{kf}; \, r\approx \frac{k r_0 f}{k_d}\approx
\frac{k_{pr}R_0}{k_d}.$$ AE is never exhausted  even when $f \to \infty$. Immortality at
infinite load is possible. Something is wrong in the model.  AE production should
decrease for large non-compensated stressors $\psi=f-r$. Let us modify the production
term in (\ref{simplest}) and add a  fitness (well-being) $W$. This fitness (well-being),
 is equal to one when the stressor load is compensated and goes to zero when the
non-compensated value of the stressor load $\psi=f-r$ becomes sufficiently large. Let us
choose the following form of $W$ for one-factor model:
\begin{equation}\label{fitness}
W(\psi)=\left(1-\frac{\psi}{\psi_0}\right), \;\; 0\leq \psi \leq \psi_0.
\end{equation}
Fitness $W(\psi)$ is a linear function on the interval $ 0\leq \psi \leq \psi_0$. It
takes its maximal value 1 at point $\psi = 0$ (completely compensated stressors) and
vanishes at $\psi=\psi_0$ (Fig.~\ref{Fittness}).

Formally, it may be continued to the whole line by constants: $W=1$ for $\psi<0$  and
$W=0$ for $\psi
> \psi_0$:
$$W(\psi)=\left(1-\frac{\psi h(\psi)}{\psi_0}\right)h\left(1-\frac{\psi
h(\psi)}{\psi_0}\right).$$

Nevertheless, it is convenient to use the simplest linear function (\ref{fitness}) and
analyse the system at the borders $\psi=0$ and $\psi=1$ separately.

The modified system of equations has the form:

\begin{equation}\label{simple}
\begin{split}
&\frac{\D r}{\D t}=-k_d r+kr_0(f-r)h(f-r);\\
&\frac{\D r_0}{\D t}=-k_{d0}r_0 - kr_0(f-r)h(f-r)+k_{pr}(R_0-r_0)W(f-r),
\end{split}
\end{equation}
where the fitness function $W(\psi)$ is given by (\ref{fitness}).

\section{Problems in definition of instant individual fitness}

We use  an individual's fitness $W$ to measure the wellbeing (or performance) of an organism. Moreover, this is an instant value, defined for every time moment. Defining of the instant measure of an individual's performance is a highly non-trivial task. The term `fitness' is widely used in mathematical biology in essentially another sense based on the averaging  of reproduction rate over a long time \citep{Haldane1932,Maynard-Smith1982,MetzNisbetGeritz1992,GorbanSelTth2007}. This is Darwinian fitness. It is non-local in time because it is the average reproduction coefficient in a series of generations and does not characterize an instant state of an individual organism.

The synthetic evolutionary approach starts with the analysis of genetic variation and studies the phenotypic effects  of that  variation  on physiology. Then it goes to the performance of organisms in the sequence of generations (with adequate analysis of the environment) and, finally, it has to return to Darwinian  fitness  \cite{Lewontin1974}. The  physiological ecologists  are focused, first of all,  on  the  observation  of  variation  in individual performance \citep{Pough1989}. In this approach we have to measure the individual performance and then link it to the Darwinian fitness.

The connection between individual performance and Darwinian fitness is not obvious. Moreover, the dependence between them is not necessarily  monotone. This observation was formalized in the theory of $r-$ and $K-$ selection \citep{MacArthurWilson1967,Pianka1970}. The terminology refers to the equation of logistic growth: $\dot{N}=rN(1-\frac{N}{K})$ ($K$  is the `carrying  capacity'  and $r$  the  maximal  intrinsic  rate  of  natural  increase). Roughly speaking, $K$ measures the  competitive abilities of individuals, and $r$ measures their  fecundity. Assuming negative correlations between $r$ and $K$, we get a question: what is better in the Darwinian sense: to increase individual competitive abilities or to increase fecundity? Earlier,   \cite{Fisher1930}  formulated a particular case of this  problem  as  follows:  ``It  would  be instructive  to know  not  only  by what
physiological  mechanism  a  just apportionment  is  made  between  the  nutriment  devoted  to  the  gonads  and that  devoted  to the  rest  of  the  parental
organism,  but  also  what  circumstances  in  the  life-history  and environment
would  render  profitable  the  diversion  of  a  greater  or lesser  share  of  the
available  resources  towards  reproduction.'' The optimal balance between individual performance and fecundity depends on environment. Thus, \cite{Dobzhansky1950} stated that in  the  tropical  zones  selection typically  favors   lower  fecundity  and  slower  development,  whereas  in  the temperate  zones  high  fecundity  and  rapid development could  increase  Darwinian fitness.

Nevertheless, the idea that the states of an organism could be linearly ordered from bad to good performance (wellbeing) is popular and useful in applied physiology. The coordinate on this scale is also called `fitness'. Several indicators are measured for fitness assessment  and then the fitness is defined as a composite of many attributes and competencies. For example, for fitness assessment in sport physiology these competencies include physical, physiological and psychomotor factors \citep{ReillyDoran2003}. The balance between various components of sport-related instant individual fitness depends upon the specific sport, age, gender, individual history and even on the role of the player  in the team (for example, for football).

\begin{figure}
\centering{
\includegraphics[width=0.55\textwidth]{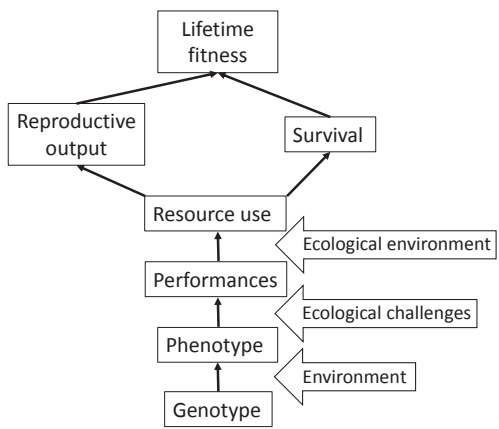}
}\caption{ Flow diagram showing the paths through from genotype to Darwinian fitness. Genotype in combination with environment determines the organismal design (the phenotype) up to some individual variations.  Phenotype determines  the limits of an individual's ability to perform day-to-day behavioral answer to main ecological challenges (performances). Performance capacity interacts with the given ecological environment and determines the resource use, which is  the key internal factor determining $r-$ and $K-$ components of fitness,
reproductive output and survival.
 \label{PerfScheme}}
\end{figure}

Similarly, the notion  `performance' in ecological physiology is `task--de\-pen\-dent' \citep{Wainwright1994} and refers to an organism's ability to carry out specific behaviors and tasks (e.g., capture prey, escape predation, obtain mates). Direct instant measurement of Darwinian fitness is impossible but it is possible to measure various instant performances several times and treat them as the components of fitness in the chain of generations. \cite{Arnold1983} proposed   several criteria for selection of the good  measure  of performance in the evolutionary study: (1) the measure  should be ecologically relevant, i.e. it measures success in the ecologically important behavior significant for survival and reproductive output; (2)  the measure  should  be phylogenetically interesting, i.e. it captures the differences between taxa and the difference between higher taxa is larger than for closed taxa, at least, for some types of performance.
The relations between performance and lifetime fitness are sketched on flow-chart (Fig.~\ref{PerfScheme}) following \cite{Wainwright1994} with minor changes. Darwinian fitness may be defined as the lifetime fitness averaged in a sequence of generations.

The idea of individual fitness is intensively used in conservation physiology \citep{WikelskiCooke2006}. An important problem is to determine how single intensive periods of stress influence individual fitness. \cite{WikelskiCooke2006} stressed that when the link between baseline physiological traits and fitness is known, conservation managers can use physiological traits as indicators to predict and anticipate future problems. Ecological success is coupled to environmental conditions via the sensitivity of physiological systems \citep{SeebacherFranklin2012}. Ideally, individual fitness is maximized when the organism can perform at a constant and optimal level despite environmental variability, but this is impossible in  the changing world for several reasons: (i) adaptation requires time and there is a lag between the changes in environment and the adaptive response, (ii) adaptation has a cost and excessive adaptation load may decrease performance because of this cost, and (iii) adaptation has its limits and even in the most plastic organisms, the capacity to compensate for environmental change is bounded.

We use the instant individual fitness (wellbeing) $W$ as a characteristic of the current state of the organism, reflecting the non-optimality of its performance: $W=1$ means the maximal achievable performance, $W=0$ means inviability (death). If the organism lives at some level of $W$ then we can consider $W$ as a factor in the lifetime fitness. Such a factorization assumes that the physiological state of the organism acts independently of other factors to determine fitness. This assumption follows the ideas of \cite{Fisher1930}. The basic assumptions of Fisher's model were analysed by \cite{Haldane1932}. `Independence' here is considered as multiplicativity, like in probability theory.  Of course, the hypothesis of independence is never absolutely correct, but it gives a good initial approximation in many areas, from data mining (na\"ive Bayes models) to statistical physics (non-correlated states).

This is the qualitative explanation of the instant individual fitness $W$. It is the most local in time level in the multiscale hierarchy of measures of fitness: instant individual fitness $to$ individual life fitness $to$ Darwinian fitness in the chain of generations. The proper language for discussion of the individual fitness gives the idea of particular performances, these are abilities of the organism to answer various specific ecological challenges. The instant individual fitness aims to combine various indicators of different performances into one quantity.

The quantitative definition of the  $W$ scale is given by its place in the equations. The change of the basic equation will cause the change of the quantitative definition. Now, we are far from the final definition of $W$. Moreover, it is plausible that for different purposes we may need different definitions of $W$.

\section{Dangerous borders}

The fitness takes the maximal value $W=1$ if the factor is fully compensated, $f=r$. Due
to equations (\ref{simple}) if $f=r$ and $r\geq 0$ then  ${\D r}/{\D t}=-k_d r \leq 0$
and ${\D W}{\D t}\leq 0$. Therefore, the fitness $W$ cannot exceed the value 1 if it is
initially below 1.

The line $W=0$ (i.e. $f-r=\psi_0$) is a {\em border of death}. If $W$ becomes negative, it means death.
On this border,
$$\mbox{ If } r_0< k_d \frac{f-\psi_0}{\psi_0} \mbox{ then } \frac{\D r}{\D t }<0\mbox{  and }\frac{\D W}{\D t }<0;$$
$$\mbox{ If } r_0> k_d \frac{f-\psi_0}{\psi_0} \mbox{ then } \frac{\D r}{\D t }>0\mbox{  and }\frac{\D W}{\D t }>0;$$

\begin{figure}
\centering{
 a)\includegraphics[height=0.3\textwidth]{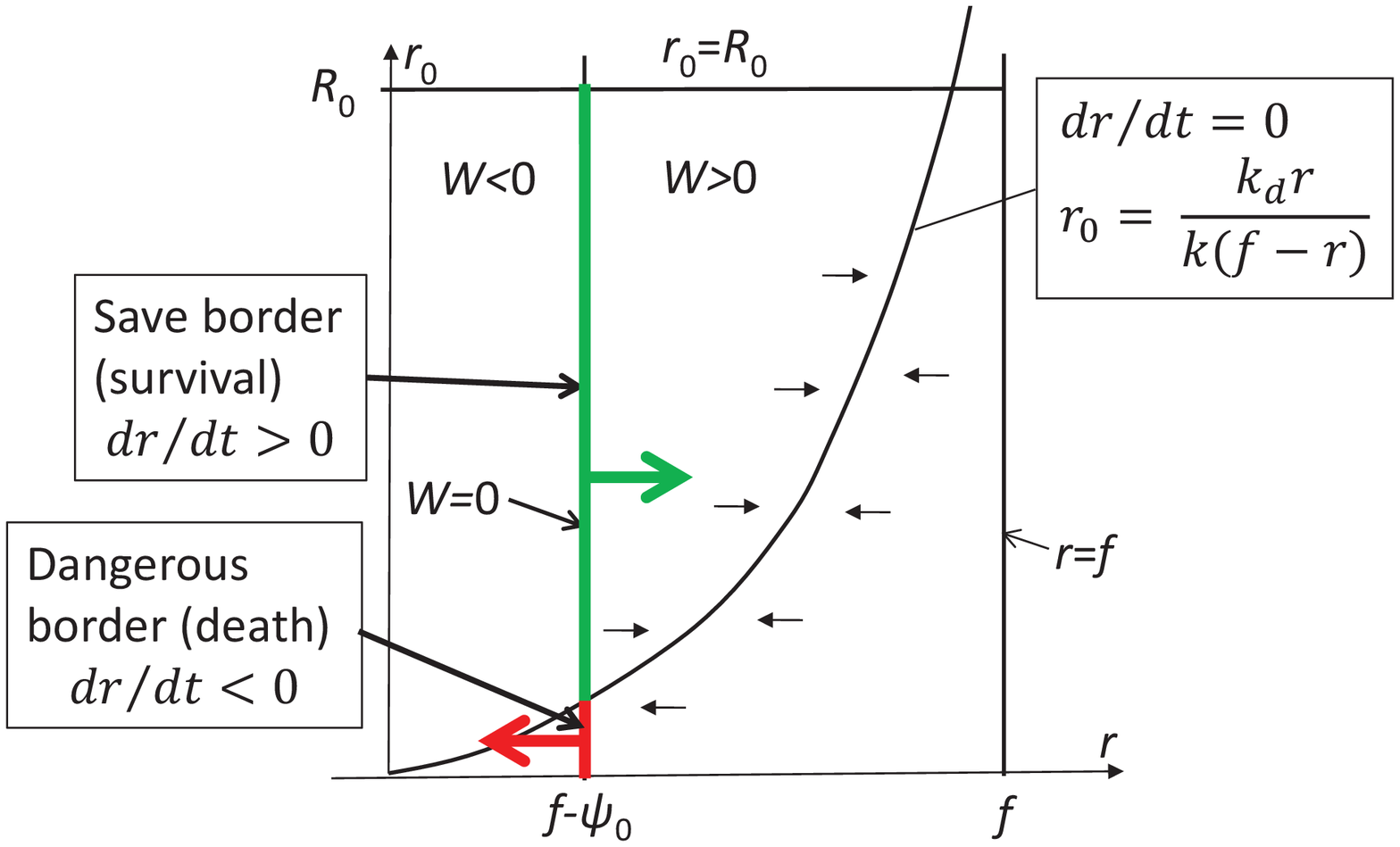} \
 b)\includegraphics[height=0.3\textwidth]{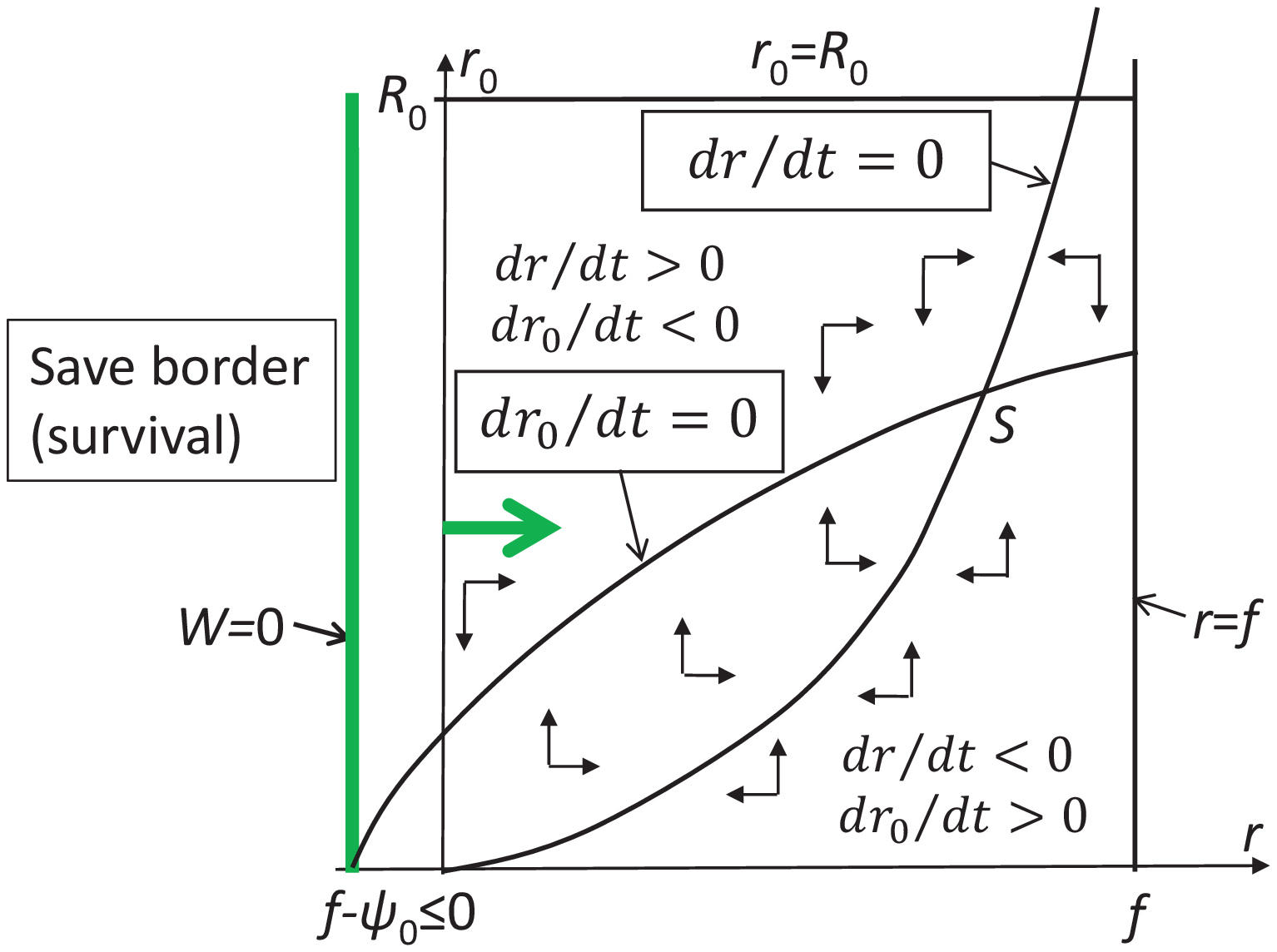}
 c)\includegraphics[height=0.3\textwidth]{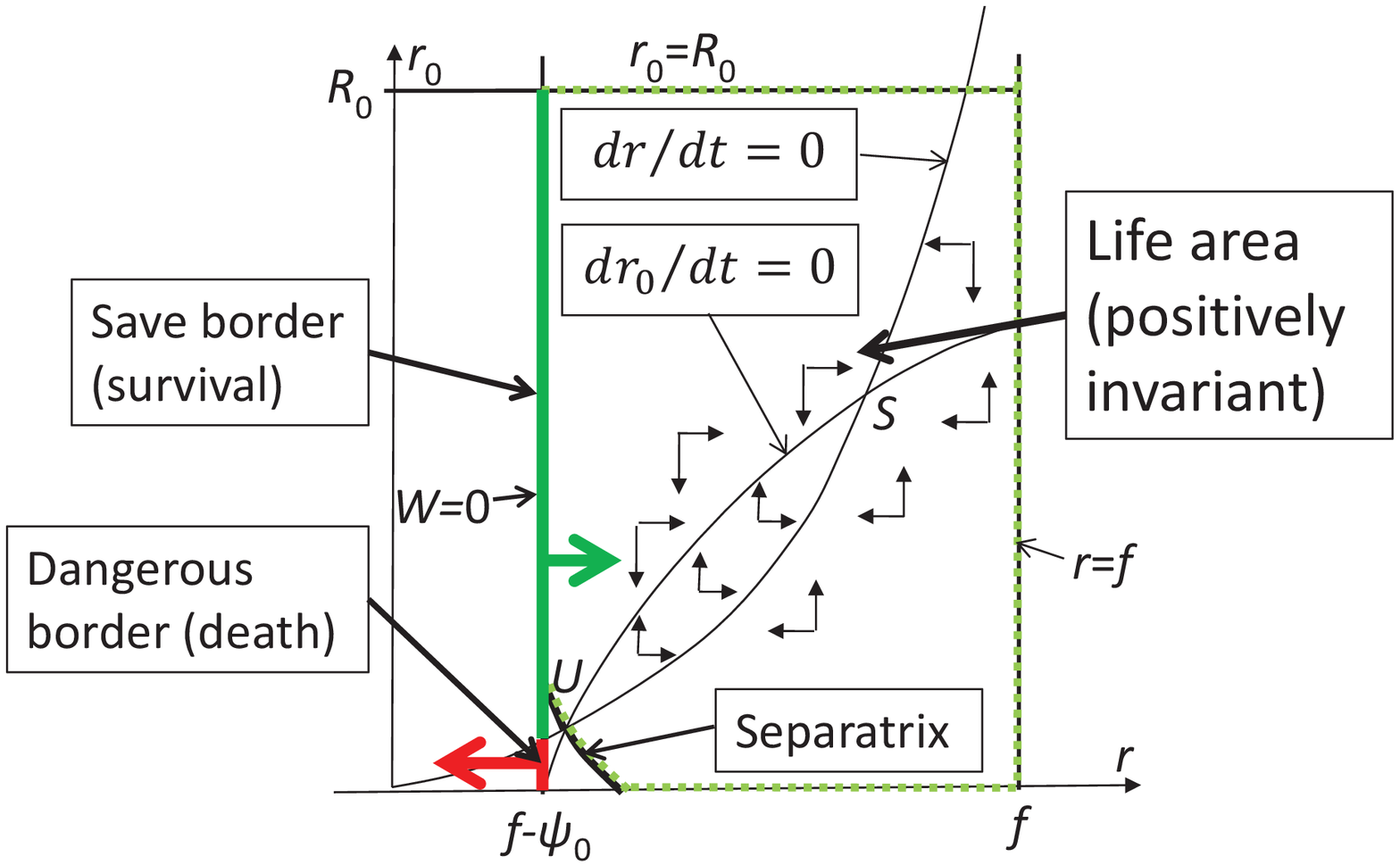} \
 d)\includegraphics[height=0.3\textwidth]{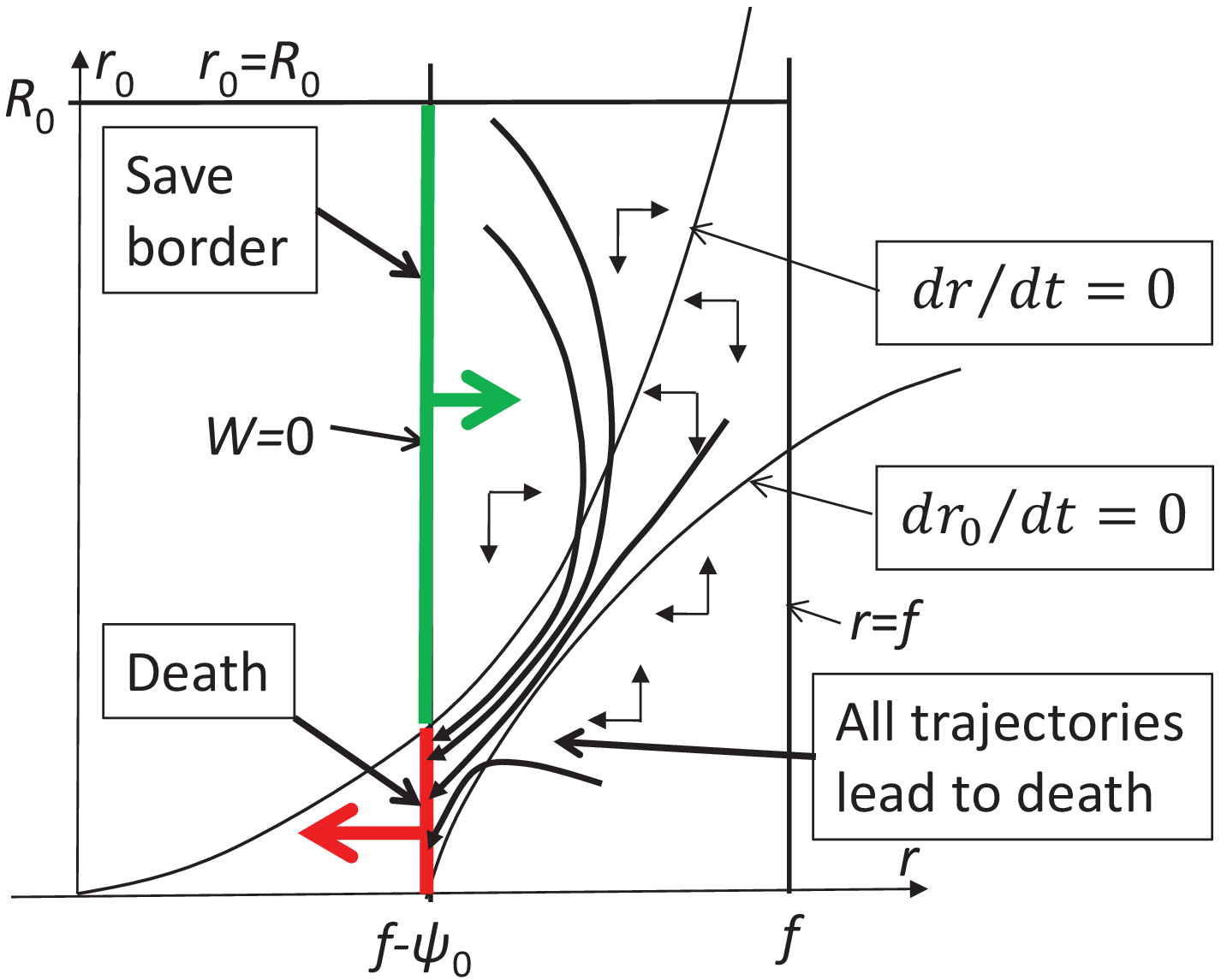}
}\caption{Safe and dangerous borders for adaptation system (\ref{simple}) for $q>0$. The
$r$-nullcline cuts the border of death $W=0$ ($r=f-\psi_0$) into two parts: $\dot{W}<0$
(dangerous border, red) and $\dot{W}>0$ (safe border, green) (a). The nullclines have in
this case (a) unique intersection point $S$ in $D$ (that is the stable equilibrium). If
$f<\psi_0$ then the whole border is safe (b). If the $r$- and $r_0$-nullclines have two
intersections, the stable ($S$) and unstable ($U$) equilibria (c), then the separatrix of
the unstable equilibrium $U$ separates the area of attraction of the dangerous border
(area of death) from the area of attraction of stable equilibrium (life area) (c). If
there exists no intersection of the nullclines in the rectangle (d) then all the
trajectories are attracting to the dangerous border. \label{Border}}
\end{figure}

The situation when $W=0$ and $\D W/\D t <0$ leads to death. Therefore, this part of the
border ($r_0< k_d (f-\psi_0)/\psi_0$) is called the {\em dangerous border}. On the
contrary, if $W=0$ but $\D W/\D t > 0$ it means survival and this border ($r_0> k_d
(f-\psi_0)/\psi_0$) is safe. The intersection point of the border of death and the
$r$-nullcline of system (\ref{simple}) separates the safe part of the border from the
dangerous part (Fig.~\ref{Border}a).

If $f\leq \psi_0$ then the whole border of death belongs to the half-plane $r\leq 0$
(Fig~\ref{Border}b). In this case, all the borders of the rectangle $D$ ($0\leq r \leq
f$, $0\leq r_0 \leq R_0$) are repulsive and the motion remains in $D$ forever, if it
starts in $D$. Below we consider the case $0 < \psi_0 < f$.  Let us analyse the system
(\ref{simple}) in the rectangle $Q$ given by the inequalities
\begin{equation}\label{Qspace}
Q: \; 0\leq r,\; f-\psi_0 \leq r \leq f, \; 0\leq r_0 \leq R_0.
\end{equation}
In the rectangle $Q$ the Heaviside functions in system (\ref{simple}) could be deleted
and this system takes a simple bilinear form
\begin{equation}\label{simpleQ}
\begin{split}
&\frac{\D r}{\D t}=-k_d r+kr_0(f-r);\\
&\frac{\D r_0}{\D t}=-k_{d0}r_0 -
kr_0(f-r)+k_{pr}(R_0-r_0)\left(1-\frac{f-r}{\psi_0}\right),
\end{split}
\end{equation}
$Q$ is not necessarily positively invariant with respect to (\ref{simpleQ}). The system
may leave $Q$ through the dangerous border.

The nullclines of this system (\ref{simpleQ}) in $Q$ are plots of monotonic functions
$r_0(r)$. The $r$-nullcline is, for $r< f$, monotonically growing convex function of $r$:
$$-k_d r+kr_0(f-r)=0, \mbox{ or } r_0=\frac{k_d r}{k
(f-r)}=\frac{k_d}{k}\left(\frac{f}{f-r}-1\right).$$

The $r_0$-nullcline is
\begin{equation*}
\begin{split}
&-k_{d0}r_0 - kr_0(f-r)+k_{pr}(R_0-r_0)\left(1-\frac{f-r}{\psi_0}\right)=0,\mbox{ or } \\
&r_0=\frac{k_{pr} R_0}{q
\psi_0}\left(1-\frac{\frac{1}{q}(k_{d0}+k\psi_0)}{r-(f-\psi_0)+\frac{1}{q}(k_{d0}+k\psi_0)}\right),
\end{split}
\end{equation*}
where $q=\frac{1}{\psi_0}k_{pr}-k\neq 0$.

The product $q\psi_0=k_{pr}-k\psi_0$ is the difference between the adaptation energy
production rate constant $k_{pr}$ and the supply coefficient $k\psi_0$ at the critical
value $f-r=\psi_0$ (the supply rate is $k(f-r)r_0$).

If $q=0$ then the $r_0$-nullcline is a straight line
$$r_0=\frac{k_{pr}
R_0}{\psi_0}\frac{r-(f-\psi_0)}{k_{d0}+k\psi_0}.$$

Geometry of the phase portraits is  schematically presented in Fig.~\ref{Border}b,c,d.
The nullclines are monotonic, the $r$-nullcline is convex, and for the case $q>0$ the
$r_0$-nullcline is concave. The area between the nullclines is positively invariant. The
phase portrait transforms from Fig.~\ref{Border}b to Fig.~\ref{Border}c and d when the
pressure of factor $f$ increases starting from safe values $f\leq \psi_0$ to high values
$f\gg \psi_0$.

\section{Resource and reserve}

Selye, Goldstone and other researchers stressed that there are different levels of the
adaptation energy supply, with lower and higher energy expenditure. \cite{Garkavi1979}
insisted that there are many levels at lower intensity of stressors, and created the
`periodic table' of the adaptation reactions. Nevertheless, we propose to formalize,
first, the two-state hypothesis.

\begin{figure}
\centering{
\includegraphics[width=0.5\textwidth]{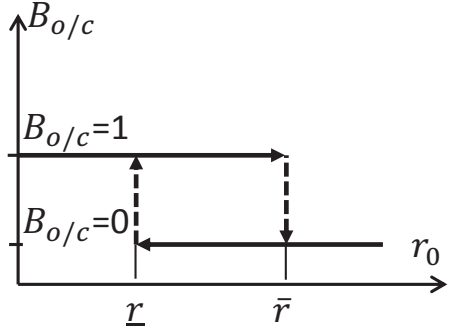}
}\caption{Resource -- reserve hysteresis. Hysteresis of reserve supply: if $B_{o/c}=0$ then
reserve is closed and if $B_{o/c}=0$  then reserve is open.  When   $r_0$ decreases and approaches
$\underline{r}$ then the supply or reserve opens (if it was closed). When
$r_0<\overline{r}$ increases and approaches $\overline{r}$ then the supply of reserve
closes (if it was open).  \label{Hysteresis}}
\end{figure}

There are two storages of AE: resource (which is always available if it is not empty) and
reserve (which becomes available when the resource becomes too low). The Boolean variable
$B_{o/c}$ describes the state of the  reserve storage: if $B_{o/c}=0$ then the reserve
storage is closed and if $B_{o/c}=1$ then the reserve storage is open. There are two
switch lines on the phase plane $(r,r_0)$: $r_0=\underline{r}$ (the lower switch line
that serves to opening the reserve storage) and $r_0=\overline{r}$ (the upper switch line
that serves to closing the reserve storage). When  the available resource $r_0$ decreases
and approaches $\underline{r}$ from above then the supply or reserve opens (if it was
closed). When the available resource $r_0<\overline{r}$ increases and approaches
$\overline{r}$ from below then the supply of reserve closes (if it was open). For
$r_0<\underline{r}$ the reserve is always open, $B_{o/c}=1$ and for $r_0>\overline{r}$
the reserve is always closed, $B_{o/c}=0$ (Fig.~\ref{Hysteresis}). These rules together
with the following equations describe the system in the rectangle $Q$ (\ref{Qspace}).

\begin{equation}\label{SysReserve}
\begin{split}
&\frac{\D r}{\D t}=-k_d r+kr_0(f-r);\\
&\frac{\D r_0}{\D t}=-k_{d0}r_0 - kr_0(f-r)+k_{rv}B_{o/c}r_{rv}(R_0-r_0)+
k_{pr}(R_0-r_0)W;\\
&\frac{\D r_{rv}}{\D t}=-k_{d1}r_{rv}-k_{rv}B_{o/c}r_{rv}(R_0-r_0)+
k_{pr1}(R_{rv}-r_{rv})W,
\end{split}
\end{equation}
where $(r,r_0)$ belongs to the rectangle $Q$ (\ref{Qspace}), $0 \leq r_{rv} \leq R_{rv}$ is the amount of reserve, $R_{rv}=const$ is the upper limit of the reserve and $W=1-\frac{f-r}{\psi_0}$ if we accept the particular simple form of fitness
function (\ref{fitness}).

For dynamics of $r_0$, the additional supply of AE from the reserve looks like the
increase of the well-being $W$ by ${k_{rv}r_{rv}}/{k_{pr}}$: after joining the last two terms
in the second equation of (\ref{SysReserve}) we get
\begin{equation}\label{SysReserveW}
\frac{\D r_0}{\D t}=-k_{d0}r_0 - kr_0(f-r)+
k_{pr}(R_0-r_0)\left(W+\frac{k_{rv}B_{o/c}r_{rv}}{k_{pr}}\right).
\end{equation}

Let us analyse the impact of reserve on the dynamics of adaptation in the small vicinity
of the border of death $W=0$. For simplicity, consider the case with sufficiently large reserve and fast reserve recovery.

\begin{figure}
\centering{
\includegraphics[width=0.85\textwidth]{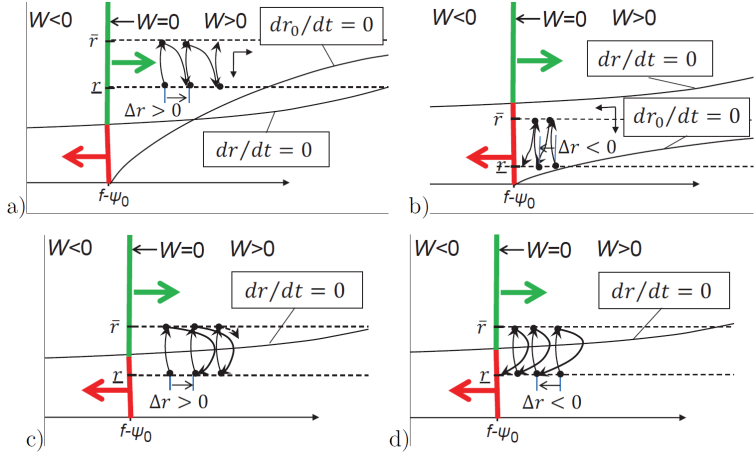}}
 \caption{Oscillating recovery (a,c) and oscillating death (b,d) near the border $W=0$ for the systems with large reserve. (Horizontally stretched sketch.)  In case (a) both $\overline{r}, \underline{r} >r^*$, in case (b) both $\overline{r}, \underline{r} <r^*$, and in cases (c),( d)  $\overline{r} >r^*> \underline{r}$, where $r^*$ is the value of $r_0$, which separates the safe border from the dangerous border on the line $W=0$ (\ref{border r}). The straight angles of possible velocities are presented for motions without research supply in cases (a) and (b) \label{OscDeath}}
\end{figure}

There are three qualitatively different cases of  the motion  in the interval $\overline{r}\geq r_0 \geq \underline{r}$  near the border $W=0$:
\begin{itemize}
\item $\overline{r}, \underline{r} >r^*$ and the motion goes above both nulclines (Fig.~\ref{OscDeath}a);
\item $\overline{r}, \underline{r} < r^*$  and the motion goes below  the $r$-nulcline but above the $r_0$-nulcline (Fig.~\ref{OscDeath}b);
\item  $\overline{r} >r^*> \underline{r}$ and the motion intersects $r$-nulcline (Fig.~\ref{OscDeath}c,d).
\end{itemize}
Here, $r^*$  is the value of $r_0$, which separates the safe border from the dangerous border on the line $W=0$,
\begin{equation}\label{border r}
r^*=\frac{k_d}{k}\frac{f-\psi_0}{\psi_0}.
\end{equation}

In all these cases the motion oscillates between the lines $r_0=\overline{r}$ and $r_0=\underline{r}$ (Fig.~\ref{OscDeath}a). When the motion
with closed reserve supply ($B_{o/c}=0$) reaches the line  $r_0=\underline{r}$ then the reserve supply switches on ($B_{o/c}=1$), the value of $r_0$ goes up fast and quickly achieves $\overline{r}$  (because of the assumption of large reserve). The value of $r$ does not change significantly during this `jump' of $r_0$ from $\underline{r}$ to $\overline{r}$. When the motion with open reserve supply  ($B_{o/c}=1$) reaches the line $r_0=\overline{r}$ then supply of reserve switches off  ($B_{o/c}=0$) and the value of $r_0$ decreases. (Note, that if the motion is sufficiently close to the border $W=0$ then it is above the nullcline of $r_0$ on the plane $(r,r_0)$, Figs.~\ref{Border}, \ref{OscDeath}.)

Consider the motion which starts on the line  $r_0=\underline{r}$  with open reserve supply. The motion returns to the same line $r_0=\underline{r}$ after the cycle: `jump up' to the line $r_0=\overline{r}$, switch reserve supply off  and `move down' without reserve supply to the line $r_0=\underline{r}$, but the value of $r$ may change. If this change $\Delta r >0$ then the system moves from the border $W=0$ (oscillating recovery, Fig.~\ref{OscDeath}a,c). If $\Delta r <0$ then the system moves to the border $W=0$ (oscillating death, Fig.~\ref{OscDeath}b,d).

\begin{figure}
\centering{
\includegraphics[width=0.85\textwidth]{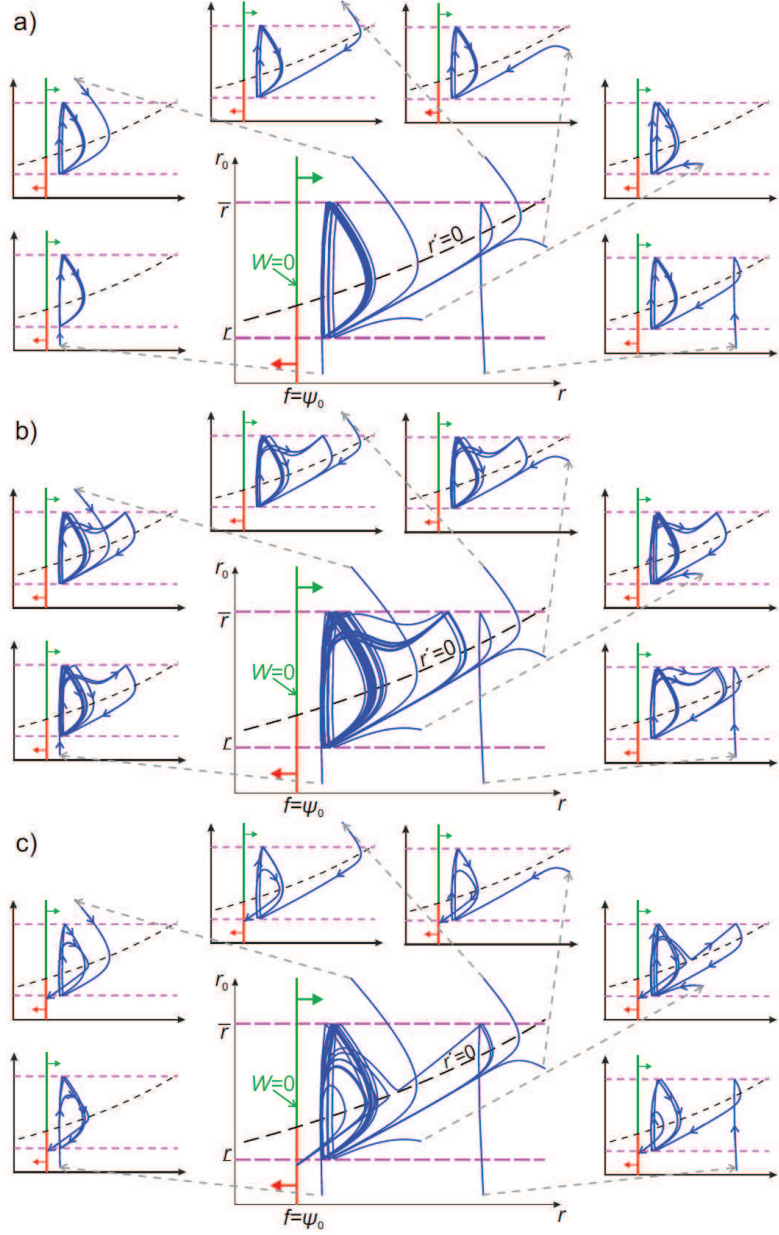}
}\caption{Oscillations near the border of death for system (\ref{SysReserve}) in projection onto the $(r,r_0)$ plane (the reserve coordinate $r_{rv}$ is hidden). For each case (a), (b), and (c) several trajectories are plotted together (central plots) and separately (side plots). At the initial points of all trajectories the reserve is full, $r_{rv}=R_{rv}$. For all cases $\overline{r}=2$, $\underline{r}=0.5$,  $R_0=10$,  $R_{rv}=5$,  $k_d=1$, $ k_{d0}=0.1$,  $k_{d1}=0.1$, $k=0.5$, $k_{pr}=2$, $k_{rv}=2$, $ \psi_0=7$, and $ f=10$. For case (a)  $ k_{pr1}= 18$ (stable oscillation), for case (b) $k_{pr1}=7$ (stable oscillations with longer orbit), for case (c) the closed orbit vanishes and the trajectories cross the  borders of death  ($k_{pr1}=3.6$). \label{cycle}}
\end{figure}

If we combine the cases Fig.~\ref{OscDeath}c (close to the border $W=0$) and d (at some distance from this border) then we can find the stable closed orbit for some combination of parameters in the limit of large reserve and fast reserve recovery. Such an orbit is presented in Fig.~\ref{cycle}a (numerical calculation). If we decrease the reserve recovering constant $k_{pr1}$ (and do not change other constants) then the closed orbit may become larger with longer time of reserve supply  (Fig.~\ref{cycle}b). The further decrease of $k_{pr1}$ leads to destruction of the closed orbit and the oscillating death appears (Fig.~\ref{cycle}d). The values of parameters were chosen just for numerical example.

Fig.~\ref{cycle}c demonstrates an important effect: the trajectories spend a long time near the places where cycles appear for different values of constants (see Fig.~\ref{cycle}a and Fig.~\ref{cycle}b) and go to the attractor (here it is death) after this delay. The delayed relaxation is a manifestation of the so--called `critical retardation': near a bifurcation with the appearance of new $\omega$-limit points, the trajectories spend a long time close to these points \citep{Gorban2004}.

The models based on Selye's idea of adaptation energy demonstrate that the oscillating
remission and oscillating death do not need exogenous reasons. These phenomena have been observed
in clinic for a long time and now attract attention in mathematical medicine and
biology. For example, \cite{Zhang2014} demonstrated recently, on a more detailed
model of adaptation in the immune system, that cycles of relapse and remission, typical
for many autoimmune diseases, arise naturally from the dynamical behavior of the system.
The notion of `oscillating remission'  is used also in psychiatry \citep{Gudayol-2015}.

\section{Distribution of adaptation energy in multifactor systems}

Usually, organisms experience a load of many factors, where  the effect of one factor could depend
on the loads of all other factors. We define a harmful factor or `stressor' as a noxious stimulus
and the `stress response' of an organism as a suite of physiological and behavioral mechanisms to cope with stress  \citep{WikelskiCooke2006}.  Revealing and description of important factors may be a non-trivial task because any biological pattern is correlated with a large number of abiotic and biotic patterns. Some of them are known, though  many are unknown. Correlations are not sufficient for extraction of main factors and the special effort and experimental study are needed to reveal causality \citep{SeebacherFranklin2012}.

The effect of action of several factors may be far from additive. There are various mechanisms of interaction between
factors in their action.  The discovery of the first non-additive interaction between
factors was done by Carl Sprengel in 1828 and Justus von Liebig in 1840 \citep{van der
Ploeg1999}. They proposed `the law of the minimum' (known also as `Liebig's law').
This law states that growth is controlled by the scarcest resource (limiting factor)
\citep{Liebig1}. It is widely known that not all systems of factors satisfy the law of the
minimum. For example, some harmful factors can  intensify effects of each
other (effect of {\em synergy} means that the harm is superadditive). The {\em colimitation} effects are also widely known
\citep{WutzlerReichstein2008}. \cite{GorbanPokSmiTyu} analysed and compared adaptation to
Liebig's and synergistic systems of factors. They formalized the idea of synergy for
multifactor systems, introduced generalized Liebig's systems and studied distribution of
AE for neutralization of the load of many factors. For this purpose, the optimality
principle was used. \cite{Tilman1980} studied resource competition. He developed an equilibrium theory based
on classification of interaction in pairs of resources. According to \cite{Tilman1980}
they may be: (1) essential, (2) hemi-essential, (3) complementary, (4) perfectly
substitutable, (5) antagonistic, or (6) switching. He also used the idea of optimality.

Evolutionary approach aims to give a universal key to the problem of optimality in
biology \citep{Haldane1932,Maynard-Smith1982,GorbanDemon1988}. The universal measure of
optimality is Darwinian fitness, that is the reproduction coefficient  averaged in a long time
\citep{GorbanSelTth2007} with some analytic simplifications, when it is possible
\citep{KarevKareva2014}, and with known generalizations for vector distributions
\citep{G1984,MetzNisbetGeritz1992}. However, there is no universal rule to measure various traits of
organisms by the changes in the average reproduction coefficient, despite exerted
efforts, development of special methods, and gaining some success
\citep{Haldane1954,WaxmanWelch2005,KingsolverPfennig2007,Shawatal2008,KarevKareva2014}. There may be
additional difficulties because the evolutionary optimality is not necessarily related to
organisms, and the non-trivial question arises: ``what is optimal?'' Another difficulty is
caused by possible non-stationarity of the optimum: selected organisms change their
environment and become non-optimal on the background of the new ecological situation
\citep{G1984}. Nevertheless, the idea of fitness is proved to be very useful. Fitness
functions are defined for different situations as intermediates between the (observable)
traits of the animal and the average reproduction coefficient.

The factors-resource models with the fitness optimization allow us to translate the elegant dynamic approach of the mathematical theory of evolution into physiological language. The key idea is to use statistical properties of physiological data instead of the data themselves. Correlations and variances are often more reliable characteristics of stress and adaptation than the values of physiological indicators \citep{GorSmiCorAd1st,GorbanSmiTyu2010,GorbanPokSmiTyu,CensiGiuliani2011,BernardiniEtAl2013}.

For formal definitions of Liebig's and synergistic systems of factors the notion of individual and instant
fitness is used. We consider organisms that are under the influence of several factors
$F_i$ with  intensities $f_i$ ($i=1,...q$). For definiteness, assume that all the factors
are harmful (this is just the sign convention plus monotonicity assumption). AE supplied
for neutralization of $i$th factor is $r_i$ and fitness $W$ is a smooth function of $q$
variables $\psi_i=f_i-r_i\geq 0$. This means that the factors are measured in the general
scale of AE units. Comparability of stressors of different nature was empirically
demonstrated and studied by \cite{SelyeAE1}. It was a strong argument for
introduction of AE. The value $f_i- r_i= 0$ is optimal (the fully compensated factor),
and any further compensation is impossible.

Assume that the vector of variables $(\psi_1,\ldots, \psi_q)$ belongs to a convex subset
$U$ of the positive orthant $\mathbb{R}^q_+$, and $W$ is defined in $U$. Harmfulness of
all factors means that
 $$\frac{\partial W(\psi_1,\ldots, \psi_q)}{\partial \psi_i}<0\mbox{ for all }i=1,\ldots,q
 \mbox{ and } (\psi_1,\ldots, \psi_q)\in U.$$

\begin{definition} A system of factors is {\it Liebig's system}, if there exists a function of one
variable $w(\psi)$ such that
\begin{equation}\label{objectiveL}
W(\psi_1,\ldots, \psi_q)=w\left(\max_{1\leq i\leq q} \{f_i-a_i r_i\}\right)\ .
\end{equation}
A system of factors is {\it anti-Liebig's system}, if there exists a function of one
variable $w(\psi)$ such that
\begin{equation}\label{objectiveA}
W(\psi_1,\ldots, \psi_q)=w\left(\min_{1\leq i\leq q} \{f_i-a_i r_i\}\right)\ .
\end{equation}
\end{definition}
In Liebig's systems fitness depends on the worst factor pressure. In anti-Liebig's
systems fitness depends on the easiest factor pressure and the factors affect the
organism only together, in strong synergy.

To generalize these polar cases of Liebig's and anti-Liebig's system, recall the notions
of {\em quasiconvex} and {\em quasiconcave} functions. A function $F$ on a convex set $U$
is {\em quasiconvex} \citep{Greenberg1971} if all its sublevel sets are convex. It means
that for every $X,Y \in U$
\begin{equation}\label{quasiconvexineq}
F(\lambda X+ (1-\lambda)Y) \leq \max\{F(X), F(Y)\} \mbox{ for all } \lambda \in [0,1]
\end{equation}
In particular, a function $F$ on a segment is quasiconvex if all its sublevel sets are
segments.

A function $F$ on a convex set $U$ is {\em quasiconcave} if $-F$ is quasiconvex. Direct
definition is as follows: A function $F$ on a convex set $U$ is {\em quasiconcave} all
its superlevel sets are convex. It means that for every $X,Y \in U$
\begin{equation}\label{quasiconcaveineq}
F(\lambda X+ (1-\lambda)Y) \geq \min\{F(X), F(Y)\} \mbox{ for all } \lambda \in [0,1]
\end{equation}
In particular, a function $F$ on a segment is quasiconcave if all its superlevel sets are
segments.

For Liebig's system the superlevel sets of $W$ are convex, therefore, $W(\psi_1,\ldots,
\psi_q)$ is quasiconcave.

For anti-Liebig's system the sublevel sets of $W$ are convex, therefore,
$W(\psi_1,\ldots, \psi_q)$ is quasiconvex.

\begin{definition} A system of factors is generalized Liebig's system
if $W(\psi_1,\ldots, \psi_q)$ is a quasiconcave function.

A system of factors is a synergistic one, if $W(\psi_1,\ldots, \psi_q)$ is a quasiconvex
function.
\end{definition}

\begin{proposition}
A system of factors is generalized Liebig's system, if and only if for any two different
vectors of factor pressures $\mathbf{\psi}=(\psi_1,... \psi_q)$ and
$\mathbf{\phi}=(\phi_1,... \phi_q)$ ($\mathbf{\psi} \neq \mathbf{\phi}$)  the value of
fitness at the average point $(\mathbf{\psi}+\mathbf{\phi})/2$ is greater, than at the
worst of points $\mathbf{\psi}$, $\mathbf{\phi}$:
\begin{equation}\label{GenLiebigEQ}
W\left(\frac{\mathbf{\psi}+\mathbf{\phi}}{2}\right) > \min
\{W(\mathbf{\psi}),W(\mathbf{\phi})\}\ .
\end{equation}
\end{proposition}

\begin{proposition}
A system of factors is a {\it synergistic} one, if for any two different vectors of
factor pressures $\mathbf{\psi}=(\psi_1,... \psi_q)$ and $\mathbf{\phi}=(\phi_1,...
\phi_q)$ ($\mathbf{\psi} \neq \mathbf{\phi}$)  the value of fitness at the average point
$(\mathbf{\psi}+\mathbf{\phi})/2$ is less, than at the best of points $\mathbf{\psi}$,
$\mathbf{\phi}$:
\begin{equation}\label{synergy}
W\left(\frac{\mathbf{\psi}+\mathbf{\phi}}{2}\right) <
\max\{W(\mathbf{\psi}),W(\mathbf{\phi})\}\ .
\end{equation}
\end{proposition}

Distribution of the supplied AE between factors should maximize the fitness function $W$
which depends on the compensated values of factors, $\psi_i=f_i-r_i$. The total amount
$r$ of the allocated AE is given:
\begin{equation}\label{Optimality}
\left\{ \begin{array}{l}
 W(f_1-r_1, f_2- r_2, ... f_q- r_q) \ \to \ \max \ ; \\
 r_i \geq 0$, $f_i- r_i \geq 0$, $\sum_{i=1}^q r_i \leq r \ .
\end{array} \right.
\end{equation}

Analysis of this optimization problem \citep{GorSmiCorAd1st,GorbanSmiTyu2010} leads to
the following statements \citep{GorbanPokSmiTyu} which sound paradoxical (if law of the
minimum is true then the adaptation makes it wrong; if law of the minimum is
significantly violated then the adaptation decreases these violations):
\begin{itemize}
\item{{\it Law of the minimum paradox}: If for a randomly selected pair, (`State of
    environment -- State of organism'), the law of the minimum is valid (everything is
    limited by the factor with the worst value) then, after adaptation, many factors (the
    maximally possible amount of them) are equally important. }
\item{{\it Law of the minimum inverse paradox}: If for a randomly selected pair, (``State
    of environment -- State of organism''), many factors are equally important  and
    superlinearly amplify each other then, after adaptation, a smaller amount of factors
    is important (everything is limited by the factors with the worst non-compensated
    values, the system approaches the law of the minimum).}
\end{itemize}

\begin{figure}
\centering{
\includegraphics[width=0.8\textwidth]{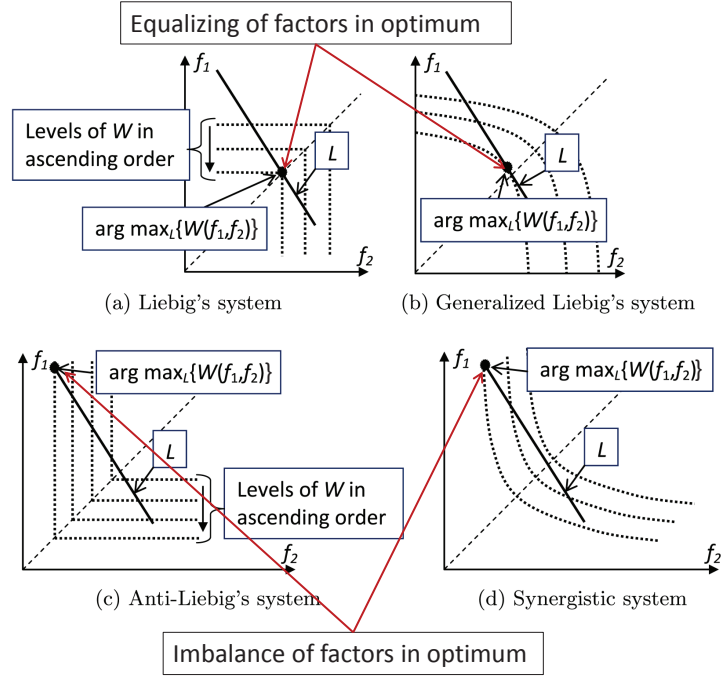}
}\caption{Distribution of AE for neutralization of several harmful factors for different
types of interactions between factors: (a) Liebig's system (the fitness $W$ depends
monotonically on the {\em maximal} non-compensated factor load only), (b) generalized
Liebig's system (the fitness $W$ is a {\em quasiconcave} function of non-compensated
factors loads), (c) anti-Liebig system (the fitness $W$ depends monotonically on the {\em
minimal} non-compensated factor load only), and (d) synergistic system (the fitness $W$
is a {\em quasiconcave} function of non-compensated factors loads). Interval $L$
represents the area of optimization. `Harmful' means that $\partial W/\partial f_i < 0$
for all factors. \label{FactorsBalance}}
\end{figure}

These properties of adaptation are illustrated by Fig.~\ref{FactorsBalance}.

Adaptation of  an organism to Liebig's system transforms  the one-dimensional picture with one
limiting factor into a high dimensional picture with many important factors. Therefore,
the well-adapted Liebig's systems should have less correlations between their attributes
than in stress. The variance (fluctuations) increases in stress. The large collection of data which supports this property of adaptation
in Liebig's system was collected since the first publication \citep{GorSmiCorAd1st} and
was reviewed by \cite{GorbanSmiTyu2010}.

Let us mention several new findings. \cite{CensiGiuliani2011} proposed using the
connectivity of correlation graphs in gene regulation networks as an indicator of analysis
of illnesses and demonstrated the validity of this approach on patients with atrial
fibrillation. \cite{BernardiniEtAl2013} studied mitochondrial network genes in the
skeletal muscle of amyotrophic lateral sclerosis patients and found correlations of gene
activities for ill patients higher than in control. \cite{KarevaEtAl2015} found signs
of this general effect in their study of consumer--resource type models and analysis of
population management strategies and their efficacy with respect to population
composition. \cite{Bezuidenhout2012} used this effect to measure the health of soil and
validated this approach.  Pareto correlation graphs, including only the highest 20\%
of correlation coefficients, were particularly useful in depicting the larger aggregated
manageability and measurability of soils. \cite{Pok2013} used analysis of dimension of
the data cloud in evaluation of human immune systems for patients with allergic disease,
either complicated or not complicated by clamidiosis. The patterns of population fluctuations are  considered as leading indicators of catastrophic shifts and extinction in deteriorating environments \citep{Dakosetal2010,DrakeGriffen2010}. The integration level in the redox
in a tissue was systematically studied \citep{Costantini2014}.

\cite{Chenetal2012} analysed microarray data of three diseases and
demonstrated that when the system reached the pre-disease state then:
\begin{enumerate}
\item There exists a group of molecules, i.e., genes or proteins, whose
average correlation coefficients of molecules
drastically increase in absolute value.
\item The average standard deviations of molecules in this group
drastically increase.
\item The average correlation coefficients of molecules between this group and any
others drastically decrease in absolute value.
\end{enumerate}
The observation 1 (increase of the correlations in the dominant group) and 2 (increase of the variance in the dominant group)  is in agreement with many of our previous results for different systems and with results of \cite{CensiGiuliani2011}, whereas the interesting observation 3 (decrease of the correlations between the dominant group and others, i.e. isolation of the dominant group) seems to be less universal (see, for example, the correlation graphs published by \cite{GorbanSmiTyu2010}).

\cite{RybRyb2012} applied the method of measurement of stress based on the Liebig's
paradox to assessing of societal stress in Ukraine. They diagnosed significant stress and
dysadaptation increase before the obvious critical events occur (the report was published
in 2012, a year before crisis). Some earlier applications to social, economical, and
financial systems were reviewed by \cite{GorbanSmiTyu2010}.

The theoretical basis of these applications can be found in the {\em quasistatic} theory of optimal resource allocation for different factors. It analyses the optimal distribution of the
total allocated AE between factors. In the previous sections of our work we develop and
analyse dynamical models of adaptation to one-factor load. We have to go ahead and create
the plausible dynamical model of adaptation to multifactor load. It is very desirable to
introduce as little new and non-measurable details as possible.

Let us start from the models (\ref{SysReserve}). First of all, we propose to use for the
total AE supply $kr_0(1-W)$ instead of $kr_0(f-r)$. For one factor with the simplest
fitness function it is just redefinition of constant $k\leftarrow k\psi_0$. Second, the
AE distribution should optimize $W$ and the simplest form of such an optimization is the
gradient descent. Immediately we get a simple system (perhaps the simplest one) which is
the direct generalization of (\ref{SysReserve}) and follows the idea of distribution of
the resource between factors for fitness increase.

\begin{equation}\label{MultiSysReserve}
\begin{split}
&\frac{\D r_i}{\D t}=-k_d r+kr_0(1-W)\frac{\frac{\partial W(\psi_1,\ldots,
\psi_q)}{\partial \psi_i}}
{\sum_i \frac{\partial W(\psi_1,\ldots, \psi_q)}{\partial \psi_i}};\\
&\frac{\D r_0}{\D t}=-k_{d0}r_0 - kr_0(1-W)+k_{rv}B_{o/c}r_{rv}(R_0-r_0)+
k_{pr}(R_0-r_0)W;\\
&\frac{\D r_{rv}}{\D t}=-k_{d1}r_{rv}-k_{rv}B_{o/c}r_{rv}(R_0-r_0)+
k_{pr1}(R_{rv}-r_{rv})W,
\end{split}
\end{equation}
where $\psi_i=f_i-r_i$;  changes of the Boolean variable $B_{o/c}$ follow the rules
formulated above (see Fig.~\ref{Hysteresis}).

The fitness function should satisfy the following requirements: it is defined in a
vicinity of $\mathbb{R}^q_+$, $0\leq W \leq 1$, $W(0)=1$, $\partial W / \partial \psi_i
\leq 0$, ${\rm grad} W = 0$ in $\mathbb{R}^q_+$ if and only if $W=1$, if $\psi_i< 0$ then
$\partial W / \partial \psi_i = 0$.

The proposed model of the adaptation to the load of many factors needs further analysis
and applications. The well-studied quasistatic model appears as a particular limiting
case of (\ref{MultiSysReserve}) for slow degradation and fast resource redistribution.

The supply of AE to neutralization of each ($i$th) factor is in (\ref{MultiSysReserve})
$$kr_0(1-W)\frac{\frac{\partial W(\psi_1,\ldots,
\psi_q)}{\partial \psi_i}} {\sum_i \frac{\partial W(\psi_1,\ldots, \psi_q)}{\partial
\psi_i}}.$$ Here, the value of the factor at $kr_0$ is always between zero and one. In
(\ref{simplest}) and (\ref{simple} we used $k_0 r_0 (f-r)$. This expression should be
corrected by saturation at large $f-r$ because the rate of AE suply cannot be arbitrarily
large: ``there is some upper limit to the amount of AE that an individual can use at any
discrete moment in time''  (Selye's Axiom 2). In (\ref{MultiSysReserve}) we get this
saturation from scratch.

\section{Conclusion and outlook}

In this paper we aim to develop a formal interpretation of Selye--Goldstone
physiological theory of adaptation energy. This is an attempt at top-down modelling
following physiological ideas. These ideas were well-prepared by their authors for
formalization and were published in the form of `axioms'.

The hierarchy of two- and three- dimensional models with hysteresis is proposed. Several
effects of adaptation dynamics are observed as oscillations in death or remission.
These oscillations do not require any external reasons and have intrinsic dynamic origin.
Observation of such effect in the clinic was already reported for some diseases.

The dynamic theory of adaptation when the organism is subject to a load of several factors needs further development.
\cite{Goldstone1952} formulated a series of questions for the future dynamical theory
of adaptation. More precisely, there was one question and several apparently
contradictory answers supported by the practical observations:

``How will one stimulus  affect an individual's power to respond to a different stimulus?
There are several different and apparently contradictory answers; yet, in different
circumstances each of these answers is probably true:
\begin{enumerate}
\item If an individual is failing to adapt to a disease he may succeed in doing so, if he
    is exposed to a totally different mild stimulus (such as slight fall of oxygen
    pressure).
\item In the process of adapting to this new stimulus he may acquire the power of
    reacting more intensely to all stimuli.
\item As a result of a severe stimulus an
    individual may not be able to adapt successfully to a second severe stimulus (such as
    a disease). If he is already adapting successfully to a disease this adaptation may
    fail when he is exposed to a second severe stimulus.
\item In some diseases (those of
    adaptation) exposure to a fresh severe stimulus may cure the disease. Exposure to an
    additional stressor will bring him nearer to death but the risk may be justifiable if
    it is likely to re-mould the adaptive mechanism to a normal form.
\end{enumerate}

Future theoretic development should help to predict, which of these contradictory answers will be
true for a given patient. Currently we are still unable to give such a prediction for individual
patients but the quasistatic theory achieves  some success in  predictions for groups and
populations \citep
{GorSmiCorAd1st,Sedov,Karmanova1996,Pokidyshevaetal1996,Svetlichnaiaetal1997,Vasi'evetal2007,RazzhevaikinShpitonkov2008,GorbanSmiTyu2010,GorbanPokSmiTyu,CensiGiuliani2011,RazzhevaikinShpitonkov2012,Bezuidenhout2012,RybRyb2012,Pok2013,BernardiniEtAl2013}.
These authors proposed and tested a {\em universal rule} to investigate in practice the amount of stress sensed by the system (and thus the danger of catastrophic changes).
The apparent universality of the top-down models of adaptation could
sometimes help in the solution of the important general problem of anticipation of
critical transitions \citep{SchefferEtal2012} and we should also try to apply  these
models in general settings.

It is necessary to validate predictions of the models. Perhaps, some further improvements are needed. For example, the classical description of the physiological reaction to  a  noxious stimulus includes three phases \citep{Selye1936}: alarm--resistance--exhaustion (the general adaptation syndrome, GAS). The alarm phase could be described more precisely than it is done in the model (\ref{SysReserve})  if we introduce an activation threshold. One Selye's axiom requires a threshold for activation of the AE supply: ``There is a threshold of AE activation   that must be present to potentiate an occupational response.'' We introduced a threshold for the  activation of reserve but did not use a threshold for the activation of the start of AE supply (thus, in our models there are two levels of AE supply). Perhaps, such a threshold of initial AE activation could help in the precise description of the alarm phase. This threshold was even included by \cite{ChrousosGold1992} in a general definition of the stress system: ``The stress system coordinates the generalized
stress response, which takes place when a stressor of any kind exceeds a threshold.''  There is some empirical evidence of the existence of a hierarchy of many activation thresholds \citep{Garkavi1998}. Construction of the models with a hierarchy of thresholds does not meet any formal difficulty but increases the number of unknown parameters.

Another improvement may be needed for the description of a dynamic response of the instant fitness to changes of factors. In the proposed models, the fitness reacts immediately. This seems to be an appropriate approximation when the intensities of the factors change slowly but in a  more general situation we have to add a differential equation for the fitness dynamics.

There also remains a theoretical (or even mathematical) challenge: the systematic and exhaustive analysis of the phase portraits of the system (\ref{SysReserve}) over the full range of parameters.

Many data about physiological, biochemical, and psychological mechanisms of adaptation and stress were collected during decades after Selye's works \citep{ChrousosGold1992,McEwen2007}. The published schemes of the stress systems and regulations include many dozens of elements. Mathematical models of important parts of homeostasis have been created \citep{PattaranitVanDenBerg2008}. In this situation, the simple models based on the AE production, distribution and spending have to prove their usefulness.

The adaptation  models introduced and analysed in this work exploit the most common phenomenological properties of  the adaptation process: homeostasis (adaptive regulation), price for adaptation (adaptation resource),  and the idea of optimization (for the
 multifactor systems). The developed models do
 not depend on the particular details of the adaptation mechanisms.

These models,  which are independent of many details, are very popular in physics,
 chemistry, ecology and many other disciplines. They aim to capture the
 main phenomena. In order to clarify the status of these models, we use the classification of models
elaborated by  \cite{Peierls1980}. He introduced six main types of models:
\begin{itemize}
\item Type 1: Hypothesis (`Could be true');
\item Type 2: Phenomenological model (`Behaves as if...');
\item Type 3: Approximation (`Something is very small, or very large');
\item Type 4: Simplification (`Omit some features for clarity') ;
\item Type 5: Instructive model (`No quantitative justification, but gives insight');
\item Type 6: Analogy (`Only some features in common');
\end{itemize}
At a first glance, we have to attribute our models
to Type 4  or even to  Type 5. Many famous models belong to these types:  the Van der  Waals model of non-perfect gases, the Debye specific heat model (Type 4);  the mean free path model for transport in gases, the Hartree-Fock model for nucleus,  and the Lotka-Volterra model of predator-prey systems (Type 5).

Nevertheless, is seems to be possible to attribute the models of adaptation elaborated in this framework of the top-down approach to the second or even to the first type. Different biological systems that have evolved can have structures with analogous forms or functions but without close common ancestor or with different intrinsic mechanisms. This is convergent evolution \citep{McGhee2011}. Some famous examples are: evolution of wings, eyes, and photosynthetic pathways.  The number of evolutionary pathways available to life may be quite limited, and the functional response to the similar environmental challenges may be similar without homology (no close common ancestor) and even with different  mechanisms.

Adaptation is a universal property of life and there are many mechanisms of adaptation. Different detailed mechanisms may produce the same  phenomenological answer at the top level because of convergent evolution.  Let us call this hypothesis the {\em Principle of phenomenological convergence}. The term `phenomenological convergence' was used in the analysis of synthetic biology by \cite{Schmidt2016} (phenomenological convergence of nature and
 technology).

The principle of phenomenological convergence results in the conclusion that the general dynamic properties of adaptation  may be much more universal than the particular biochemical and physiological mechanisms of adaptation. This manifested independence of the top phenomenological level from the bottom level (detailed mechanisms) is the result of convergent evolutions. This allows us to use AE models without solid knowledge of the intrinsic mechanism (behave as if it is true, Type 2) or even to accept them as the truth (temporarily, of course, Type 1).

\end{document}